\documentclass[10pt, conference, compsocconf]{IEEEtran}

\usepackage[english]{babel}
\usepackage[utf8x]{inputenc}

\usepackage[square,numbers]{natbib}
\ifCLASSINFOpdf
  \usepackage[pdftex]{graphicx}

\else

\fi

\usepackage[cmex10]{amsmath}
\let\oldcdot\cdot
\usepackage{breqn}
\let\cdot\oldcdot

\usepackage{algorithmic}
\usepackage{fancyhdr}

\makeatletter
\def\ps@IEEEtitlepagestyle{%
  \def\@oddfoot{\mycopyrightnotice}%
  \def\@evenfoot{}%
}
\def\mycopyrightnotice{%
  {\footnotesize Accepted to The 2017 International Conference on Computational Science and
     Computational Intelligence \hspace{2.3cm}    Copyright ©2017 IEEE\hfill}
  \gdef\mycopyrightnotice{}
}

\begin{document}
\fancypagestyle{plain}{
\fancyhf{} 
\fancyfoot[L]{978-1-4799-7492-4/15/\$31.00~\copyright2015~IEEE}
\fancyfoot[C]{}
\fancyfoot[R]{}
\renewcommand{\headrulewidth}{0pt}
\renewcommand{\footrulewidth}{0pt}
}

\pagestyle{fancy}{
\fancyhf{}
\fancyfoot[R]{}}
\renewcommand{\headrulewidth}{0pt}
\renewcommand{\footrulewidth}{0pt}
%
\title{  Solving  Minimum  \textit{k-supplier} in Adleman-Lipton model }

\author{\IEEEauthorblockN{Saeid Safaei}
\IEEEauthorblockA{Department of Computer Science  \\
University of Georgia\\
 Athens,  GA U.S.A. \\
 ssa@uga.edu}
\and
\IEEEauthorblockN{Vahid Safaei}
\IEEEauthorblockA{Mechanical Engineering\\
Yasouj University\\
 Yasouj, Iran\\}
\and
\IEEEauthorblockN{Elizabeth D. Trippe}
\IEEEauthorblockA{Institute of Bioinformatics\\
University of Georgia\\
 Athens,  GA U.S.A. \\
 elizabeth.trippe25@uga.edu}

\and
\IEEEauthorblockN{Karen Aguar}
\IEEEauthorblockA{Department of Computer Science\\
 University of Georgia\\
 Athens,  GA U.S.A. \\
kaguar@uga.edu}
\and
\IEEEauthorblockN{ Hamid R. Arabnia }
\IEEEauthorblockA{Department of Computer Science\\
 University of Georgia\\
 Athens,  GA U.S.A. \\
hra@cs.uga.edu}

}

\maketitle

\begin{abstract}
In this paper, we consider an algorithm for solving the minimum \textit{k-supplier} problem using the Adleman–Lipton model. The procedure works 
in $O(n^{2})$ steps for the minimum \textit{k-supplier} problem of an undirected graph with $n$ vertices, which is an NP-hard combinatorial optimization problem.

\end{abstract}


\IEEEpeerreviewmaketitle

\section{Introduction}
An example of non-silicon-based computing that has increased in popularity in recent years is DNA computing.  DNA computing uses DNA(deoxyribonucleic acid) strands to store information.
The DNA molecule itself has several features that make it ideal for use in computations.  The most important of these are Watson-Crick complementarity and massive parallelism. These allow
an NP-complete problem to be solved in a polynomial number of steps in contrast to a silicon-based computer, which would require an exponential number of steps.  Adleman \cite{adleman1994molecular} 
first proposed the idea of using DNA computing to solve the Hamiltonian path problem of size $n$ in $O(n)$ steps.  Lipton \cite{lipton1995dna} solved the second NP-complete problem. Subsequently, 
several scientists have solved other NP-complete problems 
\cite{ouyang1997dna,kari1998dna,dolati2008solving,safaei2009molecular,safaei2008solving,safaei2011molecular}.

To find the solution to minimum  \textit{k-supplier} NP-hard \cite{feder1988optimal} problem, we will elaborate on DNA operations defined by Adleman \cite{adleman1994molecular} 
and Lipton \cite{lipton1995dna}. The \textit{k-supplier} problem  \cite{nagarajan2013euclidean} begins with a set of clients, C, and s set of facilities, F located in a metric $(C\cup F, d) $, with a bound $k$. 
The goal of the \textit{k-supplier} problem is to find  a subset of $k$ facilities  which  minimize the maximum distance
of a client to an open facility, i.e, $ \min _{S\subseteq F:| S| =k}\max _{v \in C} d(v,S) $
where $d(v, S) = min_{u\in S} d(v, u)  $ is the minimum distance of client $v$ to
any facility in $S$. 
The graph $G$ in Fig. 1 defines such a problem. Here we define  $ \{ C=\{ 2,5 \} ,F=\{1,3,4,6 \}   \}   $ and  $k=3$  .
 For set $F$ we have 4 different subsets $ \{ S_{1}=\{1,3,4 \}, S_{2}=\{1,3,6 \}, S_{3}=\{1,4,6 \}, S_{4}=\{3,4,6 \}   \}   $ . For example the maximum distance of $S_{1}$ to Client $C$ is $ max\{ 5,2,4   \} =5  $.
 As we can see the minimum of the maximum distances of those subsets to Client  $C$  is  $ min \{ 5,6,6,6  \}=5   $.

The reminder of this paper is structured as follows.  In Section 2 , we further introduce and describe the Adleman-Lipton model and define the DNA operations used here.  In Section 3,
we introduce a DNA algorithm for solving the \textit{k-supplier} problem and we give conclusions in Section 4.

\section{ The Adleman-Lipton model}

Silicon-based computers use bytes to store information while living systems use DNA to store information in molecular form. Biomolecular systems function at the molecular level and DNA has many physical features that make it ideal to implement biological computing. 

First, DNA is a double stranded polymer in which an adenine (A) on one strand always matches thymine (T) on the other strand, so A and T are considered a Watson-Crick base pair. Also,
cytosine (C) always matches guanine (G) giving another Watson-Crick base pairing. This feature is called Watson-Crick complementarity and is one of the main physical features that makes the DNA 
molecule so unique in its structure and function. DNA strands are also described by their length in base pairs (bp) as  $l$-mers, where $l$ is the length. For example, the single strand ATTCAGCTACG 
will pair with TAAGTCGATGC to form a double stranded DNA molecule 11 bp long.

\subsection{ Operations in the Adleman-Lipton model}
According to the Adleman–Lipton model, certain operations were defined for DNA molecules. A molecule  of DNA is a finite string over the alphabet $\{A, C, G, T\}$ and a test tube is a set of molecules of DNA.

(1)
Merge$(T_{1},T_{2})$:  It accepts two test tubes, $T_{1}$ and $T_{2}$, and it finds $T_{1}\cup T_{2}$ which is stored in $T_{1}$  while $T_{2}$ is left empty \cite{xiao2005solving}.

(2)
Detect$(T)$: It accepts a test tube $T$, and it outputs “no” if $T$ is empty, otherwise, it outputs “yes”\cite{wang2006dna}.

(3)
Separation$(T_{1},X,T_{2})$: It accepts a test tube, $T_{1}$, and a single string, $X$. It selects all single strands containing $X$ from $T_{1}$, and generates a test tube, $T_{2}$, with the selected strands \cite{xiao2005solving}.

(4)
Selection$(T_{1},L,T_{2})$: It accepts a test tube, $T_{1}$, and a given integer, $L$. It selects all strands with length $L$ from $T_{1}$, and generates a test tube, $T_{2}$, with the selected strands \cite{xiao2005solving}.

(5)
Annealing $(T)$: It accepts a test tube, $T$, and the output is all feasible double strands in $T$ \cite{xiao2005solving}.

(6)
Denaturation $(T)$: It accepts a test tube, $T$, and it separates each double strand in $T$ into two single strands \cite{xiao2005solving}.

(7)
Discard $(T)$: It accepts a test tube, $T$, and it removes $T$ \cite{li2015dna}.

(8)
Append$(T,Z)$: It accepts a test tube, $T$, and a short singled stranded DNA fragment, $Z$. This fragment, $Z$, is appended onto the end of every strand in the tube $T$ \cite{li2015dna,matsakis2010solving}.

Because the previous operations have a constant number of biological steps, the complexity of each operation is assumed to be $O(1)$. 


\section{  DNA algorithm for the minimum \textit{k-supplier} problem }
Let $G=(V,E)$ be a edge-weighted  graph with the set of vertices  $ V=\{ V_{i}|i=1,2, \ldots ,n\} $, the set of edges $ E=\{ e_{i,j} | 1 \leq i,j\leq n,i\neq j\}  $ 
and $ SP=\{ sp_{i,j} | 1 \leq i,j\leq n,i\neq j\}  $.  Note that  $ e_{i,j} $ is in E if the
vertices $v_{i}$ and $v_{j}$ are connected by an edge.  Also, $sp_{i,j}$ shows the weight of the shortest path from  $v_{i}$ to $v_{j}$.
We have

$| E| \leq \dfrac {1} {2}n( n+1)$.

In the following, we use the symbols $ \#,0,1,2,X,A_{i},B_{i}(i=1,2…,n)$ and $ w_{i,j} $ to denote distinct DNA single strands for which $||\#||=||A_{i}||=||B_{i}||=||0||=||1||=||2||=||X||$, 
where $|| . ||$ denotes the length of the DNA single strand. The symbols $ A_{i}B_{i} (i=1,2,…,n) $ denote the vertex $ v_{i} $. The symbol $\#$ is the signal of the start and the signal of the end.
Suppose that all weights in the given graph are commensurable. The DNA single strands $ w_{i,j} $ are used to
denote the weights of the shortest path  $ s_{i,j} $ from $ v_{i} $ to $v_{j}$ with $ ||w_{i,j}||=sp_{i,j}* w $ where $w$ is a constant, e.g., take $ w=10 $ mer in the following
discussion, Then, the $ ||w_{i,j}||=10* sp_{i,j} $. Let $ m = max_{sp_{i,j} ∈  SP } \ w_{i,j} $ and we consider  $||\#||=||A_{i}||=||B_{i}||=||0||=||1||=||2||=||X||=10 $.  
Since the length of $||w_{i,j}||=10* sp_{i,j} $ then we can show $w_{i,j}$ by the combination of single strand $X$ with itself
so $w_{i,j}=\underbrace{XX...XX}_{\text{$sp_{i,j}$ }}$.
Let

$ \\ P=\{ 0,1,2, \# A_{1},B_{d}A_{d+1}, B_{n}\# ,   w_{i,j} |e_{i,j}\in E ,  d=1,2 ,..., n-1 \}  $

$ \\ Q=\{ \overline {\#},\overline {A_{d}0B_{d}} , \overline {A_{d}1B_{d}} , \overline {A_{d}2B_{d}}  |e_{i,j}\in E , 1 \leq i,j\leq n ,d=1,2,...,n-1 \}  $

$ \overline {}. \\ $
We designed the following algorithm to solve the minimum \textit{k-supplier} problem, which we describe below.


\subsection{ Produce all possible subsets of $E$ }

\begin{figure}
\centering
\includegraphics[width=0.5\textwidth]{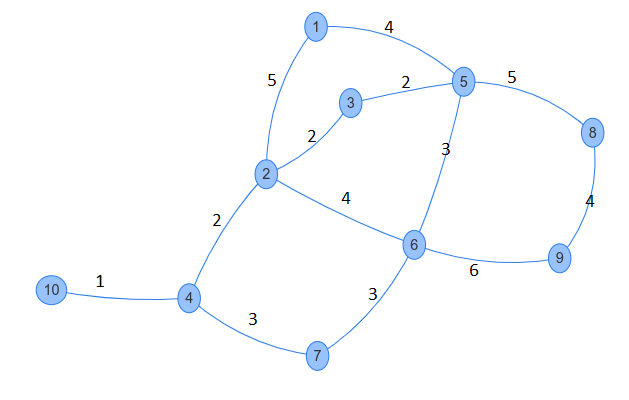}
\caption{\label{fig:Graph} $ \{ C=\{ 2,5 \} ,F=\{1,3,4,6 \}   \}   $ and  k=3 }
\end{figure}

In this step, we consider three subsets, $S$, $C$, $R$, and we want to produce all states. We can put
all vertices in these subsets.  $S$, $C$, and $R$ are subsets which show open facility, Client, and 
renaming vertices respectively. \\

$ (1-1) Merge(P,Q);$

$ (1-2) Annealing(P);$

$ (1-3) Denaturation(P);$

$ (1-4) Separation(P,\# A_{1}B_{1},T_{tmp});$

$ (1-5) Discard(P);$

$ (1-6) Separation(T_{tmp},{ A_{1}B_{1}\#},P).$ \\

After the above six steps of manipulation, the single strands in tube $P$ will encode all subsets. For example, the single strands 
$A_{l}0B_{l},A_{l}1B_{l},A_{l}2B_{l} \ | \ 1 \leq  l \leq n   $ show vertex $l$ belongs to $C$, $S$, $R$ respectively.
For another example, from the graph in Fig. 1, we have the single strand \\
$\#A_{1}1B_{1}A_{2}0B_{2}A_{3}1B_{3}A_{4}1B_{4}A_{5}0B_{5}A_{6}2B_{6}A_{7}1B_{7}A_{8}2B_{8}$ \\ $A_{9}2B_{9}A_{10}2B_{10}\# ∈ P $ which corresponds to the subsets  $ \{ C=\{ 2,5 \} ,S=\{ 1,3,4,7 \} ,R=\{ 6,8,9,10 \}  \}  $.
As we can see, it is an invalid subset, because in  Fig 1 we defined $ C=\{ 2,5 \} $ and
$ F=\{ 1,3,4,6 \} $ and $S$ should be a subset of $F$. Also, the cardinality of $S$  must be 3. 

Here is another example.  \\
$\#A_{1}1B_{1}A_{2}0B_{2}A_{3}1B_{3}A_{4}2B_{4}A_{5}0B_{5}A_{6}1B_{6}A_{7}2B_{7}A_{8}2B_{8}$ \\ $A_{9}2B_{9}A_{10}2B_{10}\#∈P $  corresponds to the subset
$ \{ C=\{ 2,5 \} ,S=\{ 1,3,6 \} ,R=\{ 4,7,8,9,10 \}  \} $. This is a valid subset according to Fig 1. This algorithm is finished in $O(1)$  since we have six steps.


\subsection{ Separate valid subsets }

As we mentioned before, \\
$\#A_{1}1B_{1}A_{2}0B_{2}A_{3}1B_{3}A_{4}1B_{4}A_{5}0B_{5}A_{6}2B_{6}A_{7}1B_{7}A_{8}2B_{8}$ \\ $A_{9}2B_{9}A_{10}2B_{10}\# ∈ P $
is invalid subset. In this step we want to remove those invalid subsets. 
If $ v_{l} \in C$, then all valid strands should contain $A_{l}0B_{l}$. So all  strands which contains $A_{l}1B_{l}$ or $A_{l}2B_{l}$  are  invalid because $A_{l}1B_{l}$ or $A_{l}2B_{l}$  denotes 
 $ v_{l} \in S $ or $ R $. We repeat this process for $ v_{l} \in F$. The algorithm is as follows:

$For \ j=1 \ to \ j=n $

{\qquad $IF \ V_{j} \in C$ Then }

{\qquad\qquad$(2-1) Separation(P,A_{j}1B_{j},T_{1})  $ }

{\qquad\qquad$(2-2) Separation(P,A_{j}2B_{j},T_{2})  $ }

{\qquad\qquad$(2-3) Discard(T_{1})$}

{\qquad\qquad$(2-4) Discard(T_{2})$}

{\qquad$End \ IF$ }

{\qquad$ IF \ V_{j} \in \ F \ Then $ }

{\qquad\qquad$(2-5) Separation(P,A_{j}0B_{j},T_{1})  $ }

{\qquad\qquad$(2-6) Separation(P,A_{j}2B_{j},T_{2})  $ }

{\qquad\qquad$(2-7) Discard(T_{1})$}

{\qquad\qquad$(2-8) Discard(T_{2})$}

{\qquad$End \ IF$ }

$End \ For$

This algorithm is finished in $O(n)$ steps since each manipulation above works in $ O(1) $ steps.


\subsection{ Count the number of members in each subset }
  
In previous algorithm, we produced single strands for each subset. If a single strand contains $B_{k}1A_{k} $, it means $v_{k} $ belongs to  subset 
S and we will add the single string $X$ to the end of that strand. the number of string $X$ at the end of each strands shows the cardinality of S.

At the end we will count the number of vertices belong to each specific subset.

In this algorithm, we need those subsets which contains exactly $K$ members. 
First we choose all subsets which contains $K+1$ members and remove them. From those remaining, we choose all subsets contain $K$ members.
  
$For \ j=n \ to \ j=1 $

{\qquad $ IF \ V_{j} \in F \ Then $ }

{\qquad\qquad$(3-1) Separation(P,w_{i,j},T_{1})  $ }

{\qquad\qquad$(3-2) Append(T_{1},w_{i,j})$}

{\qquad\qquad$(3-4) Merge(P,T_{1})$}

{\qquad$ End  \ IF $ }

$End \ For$

We consider the previous example, \\
$\#A_{1}1B_{1}A_{2}0B_{2}A_{3}1B_{3}A_{4}2B_{4}A_{5}0B_{5}A_{6}1B_{6}A_{7}2B_{7}A_{8}2B_{8}$ \\ $A_{9}2B_{9}A_{10}2B_{10}\# $. 
This algorithm add $XXX$ to the end of this strand. In this case we have \\
$\#A_{1}1B_{1}A_{2}0B_{2}A_{3}1B_{3}A_{4}2B_{4}A_{5}0B_{5}A_{6}1B_{6}A_{7}2B_{7}A_{8}2B_{8}$ \\ $A_{9}2B_{9}A_{10}2B_{10}\#XXX $. 

This algorithm is finished in $O(n)$ steps since each manipulation above works in $ O(1) $ steps.


  \subsection{ Find the longest distance of each subset to  the open facility}
For each path we have a DNA strand. In previous algorithms, we counted the number of all vertices in the subset.
 In the definition of the problem we defined  $d(v, S) = min_{u\in S} d(v, u)  $ where the goal is to open a subset of $k$ facilities so as to minimize the maximum distance
of a client to an open facility. To achieve this purpose we will create a descending array for all paths.
The first element of this array is the longest one. For example, the first element is $w_{l,m}$. As we mentioned before, we divided all vertices into three subsets,
$C$, $S$, $R$, we need to collect all subsets where $v_{l}$ belongs to C and $V_{m}$ belongs to $S$ or vice versa.
This means we found all subsets for which the longest path of client to open facility is $w_{l,m}$ and we put those subsets into a new set. For the next step, 
we choose the second longest path and repeat the same process which we will continue for all path.
To run the algorithm, we create a two-dimensional array based on descending order of paths. The first element is the source of the path and the second element is the destination of that path.
For example in Fig 1, the longest path from an element in $C$ to an element in $F$ is a path from $v_{5}$ to  $v_{4}$ which is equal to 6.
In the first step, we select all strands which contain $B_{5}0A_{5}$ which is equivalent to $ v_{5} \in C $ and $B_{4}1A_{4}$ which is equivalent to $ v_{4} \in F$. We append   $w_{5,4}$ to the end of those strands. \\

For $i=1$ to $i=n^{2} $

{\qquad $(4-1) Separation(P,B_{arr[i][1]}1A_{arr[i][1]},T_{1})  $ }

{\qquad $(4-2) Separation(T_{1} ,B_{arr[i][2]}0A_{arr[i][2]},T_{2})           $ }

{\qquad $(4-1) Separation(P,B_{arr[i][1]}0A_{arr[i][1]},T_{3})  $ }

{\qquad $(4-2) Separation(T_{3} ,B_{arr[i][2]}1A_{arr[i][2]},T_{4})           $ }

{\qquad$(4-4) Merge(T_{2},T_{4})$}

{\qquad$(4-5) Append(T_{2},w_{i,j})$}

{\qquad$(4-6) Merge(T_{5},T_{2})$}

{\qquad$(4-7) Merge(P,T_{1})$}

$End \ For$

$(4-8) Discard (P)$

$(4-9) Merge(P,T_{2})$

This algorithm is finished in $ O(n^{2}) $ steps since each manipulation above works in $ O(1) $ steps.


\subsection{ Find the optimal solution }
 We need to find a subset with minimum distance.  We will introduce two different algorithms both of which run in $O(n^{2})$.
 
 \subsubsection{ First algorithm }
 The output of previous algorithm , is a set of strands with this format \\
  $\#A_{1}?B_{1}A_{2}?B_{2}A_{3}?B_{3}A_{4}?B_{4}A_{5}?B_{5}A_{6}?B_{6}A_{7}?B_{7}A_{8}?B_{8}$ \\ $A_{9}?B_{9}A_{1?}?B_{1?} \# W_{i,j} | \  1 \leq i,j\leq n , $
 $ ? \in \{0,1,2\}$. Since we considered $ ||\#||=||A_{i}||=||0||=||1||=||2||=||B_{i}||=10 $ then the length of each strand is 
$ ||\#||+||A_{1}||+||1||+||B_{1}|| +...+||A_{n}||+||1||+||B_{n}||+||\#||+||w_{i,j}|| $.
So the length of each strand is $ 30*n+||\#||+||\#||+||w_{i,j}||= 30*n+20+||w_{i,j}||$.
Since the lengths of the first parts of each strand are identical, the shortest strand is the one which has the shortest $||w_{i,j}|| \ |  \ 0 \leq i,j \leq n$.

Suppose  $ u_{i,j} $ are used to
denote the weights of edge  $ e_{i,j} $ . Let $ Y = max_{e_{i,j} ∈  E } \ u_{i,j} $. Then the length of each shortest path from two arbitrary vertices is smaller than 
$Y*n^{2}$

$For  \ i=1 \ to i=Y*n^{2} \  $

{\qquad$(5.1-1) \ Selection(P,30n+20+10*i,T)$}

{\qquad$(5.1-2) \ IF \ Detect (T) $ is “yes”, then  Exit}

$End \ For $

This algorithm is finished in $ O(n^{2}) $ steps since each Y is a constant.

 \subsubsection{ Second  algorithm }
 As we discussed, we created  $||w_{i,j}|| \ | \ 0 \leq i,j \leq n$ strand with the combination of single strand X.
 since the actual format of each strand is \\
 $\#A_{1}?B_{1}A_{2}?B_{2}A_{3}?B_{3}A_{4}?B_{4}A_{5}?B_{5}A_{6}?B_{6}A_{7}?B_{7}A_{8}?B_{8}$ \\ $A_{9}?  B_{9}A_{10}?B_{10}\#  \underbrace{XX...XX}_{\text{$sp_{i,j}$ }}
 |  1 \leq i,j\leq n , ? \in \{0,1,2\}$.
 the shortest  $||w_{i,j}|| $ is X and the longest one could be 
 $\underbrace{XX...XX}_{\text{$Y*n^{2}/10$ }}$ . so we are looking for strands which contains X. if exist then exit .
 else we will looking for XX . and continue this process to find minimum solution.
 
 $For \ i=1 \ to \ i=Y*n^{2} \  $
 
{\qquad $ (5.2-1) Separation (P,\underbrace{X...X}_{\text{$ i \ times $ }},T_{1})$ }
 
{\qquad $ (5.2-2) Discard(P)   $ }
 
{\qquad $ (5.2-3) Separation (T_{2},\underbrace{X...X}_{\text{$ i+1 \ times $ }},T_{2})$ }
 
{\qquad $ (5.2-4) Merge (P,T_{1} ) $ }
 
{\qquad$(5.2-5) \ IF \ Detect (T) $ is “yes”, then  Exit }
 
$End \ For $

This algorithm is finished in $ O(n^{2}) $ steps since each Y is a constant.


\section{ Conclusions}

Building on Adleman's and Lipton's work on DNA computing,  we have solved the minimum \textit{k-supplier} problem which is an NP-hard problem in $O(n^{2})$.  The \textit{k-supplier} problem 
is a combinatorial optimization problem which has applications in finding locations for warehouses, clustering data, etc. In this algorithm, we first produced a space which contained 
all possible solutions and removed the invalid solutions. In the last part, we searched the solution space to find the minimum value.  The advantage of DNA computing is that we can generate 
a solution space in $O(1)$, which is exponential in silicon-based computers. Since it is hard to replace mathematical operations with biological ones and Cook's theorem is not always valid for
DNA computing \cite{guo2005optimal}, our work has value by describing a new algorithm for one NP-hard problem. Also it is valuable to introduce new algorithms for problems in Mechanical engineering    \cite{aghniaey2014exergy,aghniaey2014comparison}   , Civil engineering \cite{chorzepadesign,chorzepa2016hurricane}  and forecasting problems  \cite{fazli2013comparative,fazli2013designing}.

\bibliographystyle{IEEEtran}
\bibliography{IEEEabrv,ref1.bib}

\begin{thebibliography}{21}
\providecommand{\natexlab}[1]{#1}
\providecommand{\url}[1]{#1}
\csname url@samestyle\endcsname
\providecommand{\newblock}{\relax}
\providecommand{\bibinfo}[2]{#2}
\providecommand{\BIBentrySTDinterwordspacing}{\spaceskip=0pt\relax}
\providecommand{\BIBentryALTinterwordstretchfactor}{4}
\providecommand{\BIBentryALTinterwordspacing}{\spaceskip=\fontdimen2\font plus
\BIBentryALTinterwordstretchfactor\fontdimen3\font minus
  \fontdimen4\font\relax}
\providecommand{\BIBforeignlanguage}[2]{{%
\expandafter\ifx\csname l@#1\endcsname\relax
\typeout{** WARNING: IEEEtranN.bst: No hyphenation pattern has been}%
\typeout{** loaded for the language `#1'. Using the pattern for}%
\typeout{** the default language instead.}%
\else
\language=\csname l@#1\endcsname
\fi
#2}}
\providecommand{\BIBdecl}{\relax}
\BIBdecl

\bibitem[Adleman(1994)]{adleman1994molecular}
L.~M. Adleman, ``Molecular computation of solutions to combinatorial
  problems,'' \emph{Nature}, vol. 369, p.~40, 1994.

\bibitem[Lipton(1995)]{lipton1995dna}
R.~J. Lipton, ``Dna solution of hard computational problems,'' \emph{Science},
  vol. 268, no. 5210, p. 542, 1995.

\bibitem[Ouyang et~al.(1997)Ouyang, Kaplan, Liu, and Libchaber]{ouyang1997dna}
Q.~Ouyang, P.~D. Kaplan, S.~Liu, and A.~Libchaber, ``Dna solution of the
  maximal clique problem,'' \emph{Science}, vol. 278, no. 5337, pp. 446--449,
  1997.

\bibitem[Kari et~al.(1998)Kari, P{\u{a}}un, Rozenberg, Salomaa, and
  Yu]{kari1998dna}
L.~Kari, G.~P{\u{a}}un, G.~Rozenberg, A.~Salomaa, and S.~Yu, ``Dna computing,
  sticker systems, and universality,'' \emph{Acta Informatica}, vol.~35, no.~5,
  pp. 401--420, 1998.

\bibitem[Dolati et~al.(2008)Dolati, Haghighat, Safaei, and
  Mozaffar]{dolati2008solving}
A.~Dolati, M.~S. Haghighat, S.~Safaei, and H.~Mozaffar, ``Solving minimum
  beta-vertex separator problems in the adleman-lipton model.'' in \emph{FCS},
  2008, pp. 97--101.

\bibitem[Safaei et~al.(2009)Safaei, Dalvand, Esmaeili, and
  Safaei]{safaei2009molecular}
S.~Safaei, B.~Dalvand, B.~Esmaeili, and V.~Safaei, ``Molecular solutions for
  the minimum edge dominating set problem on dna-based supercomputing.'' in
  \emph{FCS}, 2009, pp. 32--36.

\bibitem[Safaei et~al.(2008)Safaei, Mozaffar, and Esmaeili]{safaei2008solving}
S.~Safaei, H.~Mozaffar, and B.~Esmaeili, ``Solving minimum k-center problem in
  the adleman-lipton model.'' in \emph{FCS}, vol.~8, 2008, pp. 182--186.

\bibitem[Safaei et~al.(2011)Safaei, Dalvand, Safaei, and
  Safaei]{safaei2011molecular}
N.~Safaei, B.~Dalvand, S.~Safaei, and V.~Safaei, ``Molecular solutions for the
  maximum k-facility dispersion problem on dna-based supercomputing.''\hskip
  1em plus 0.5em minus 0.4em\relax FCS, 2011.

\bibitem[Feder and Greene(1988)]{feder1988optimal}
T.~Feder and D.~Greene, ``Optimal algorithms for approximate clustering,'' in
  \emph{Proceedings of the twentieth annual ACM symposium on Theory of
  computing}.\hskip 1em plus 0.5em minus 0.4em\relax ACM, 1988, pp. 434--444.

\bibitem[Nagarajan et~al.(2013)Nagarajan, Schieber, and
  Shachnai]{nagarajan2013euclidean}
V.~Nagarajan, B.~Schieber, and H.~Shachnai, ``The euclidean k-supplier
  problem,'' in \emph{International Conference on Integer Programming and
  Combinatorial Optimization}.\hskip 1em plus 0.5em minus 0.4em\relax Springer,
  2013, pp. 290--301.

\bibitem[Xiao et~al.(2005)Xiao, Li, Zhang, and He]{xiao2005solving}
D.~Xiao, W.~Li, Z.~Zhang, and L.~He, ``Solving maximum cut problems in the
  adleman--lipton model,'' \emph{BioSystems}, vol.~82, no.~3, pp. 203--207,
  2005.

\bibitem[Wang et~al.(2006)Wang, Xiao, Li, and He]{wang2006dna}
Z.~Wang, D.~Xiao, W.~Li, and L.~He, ``A dna procedure for solving the shortest
  path problem,'' \emph{Applied mathematics and computation}, vol. 183, no.~1,
  pp. 79--84, 2006.

\bibitem[Li et~al.(2015)Li, Patrikeev, and Xiao]{li2015dna}
W.~Li, E.~Patrikeev, and D.~Xiao, ``A dna algorithm for the maximal matching
  problem,'' \emph{Automation and Remote Control}, vol.~76, no.~10, pp.
  1797--1802, 2015.

\bibitem[Matsakis(2010)]{matsakis2010solving}
N.~Matsakis, ``Solving the rural postman problem using the adleman-lipton
  model,'' \emph{arXiv preprint arXiv:1012.2527}, 2010.

\bibitem[Guo et~al.(2005)Guo, Chang, Ho, Lu, and Cao]{guo2005optimal}
M.~Guo, W.-L. Chang, M.~Ho, J.~Lu, and J.~Cao, ``Is optimal solution of every
  np-complete or np-hard problem determined from its characteristic for
  dna-based computing,'' \emph{BioSystems}, vol.~80, no.~1, pp. 71--82, 2005.

\bibitem[Aghniaey and Mahmoudi(2014)]{aghniaey2014exergy}
S.~Aghniaey and S.~M.~S. Mahmoudi, ``Exergy analysis of a novel absorption
  refrigeration cycle with expander and compressor,'' \emph{Indian Journal of
  Scientific Research}, vol.~1, pp. 815--822, 2014.

\bibitem[Aghniaey et~al.(2014)Aghniaey, Mahmoudi, and
  Khalilzad-Sharghi]{aghniaey2014comparison}
S.~Aghniaey, S.~M.~S. Mahmoudi, and V.~Khalilzad-Sharghi, ``A comparison
  between the novel absorption refrigeration cycle and the conventional
  ammonia-water absorption refrigeration cycle.''\hskip 1em plus 0.5em minus
  0.4em\relax International Conference on Heat Transfer, Fluid Mechanics and
  Thermodynamics, 2014.

\bibitem[Chorzepa et~al.()Chorzepa, Saeidpour, Christian, and
  Durham]{chorzepadesign}
M.~G. Chorzepa, A.~Saeidpour, J.~Christian, and S.~Durham, ``Design of coastal
  bridges against severe storms and sea-level rise.''

\bibitem[Chorzepa et~al.(2016)Chorzepa, Saeidpour, Christian, and
  Durham]{chorzepa2016hurricane}
M.~Chorzepa, A.~Saeidpour, J.~Christian, and S.~Durham, ``Hurricane
  vulnerability of coastal bridges using multiple environmental parameters,''
  \emph{International Journal of Safety and Security Engineering}, vol.~6,
  no.~1, pp. 10--18, 2016.

\bibitem[Fazli and Lebraty(2013)]{fazli2013comparative}
M.~S. Fazli and J.-F. Lebraty, ``A comparative study on forecasting polyester
  chips prices for 15 days, using different hybrid intelligent systems,'' in
  \emph{Neural Networks (IJCNN), The 2013 International Joint Conference
  on}.\hskip 1em plus 0.5em minus 0.4em\relax IEEE, 2013, pp. 1--7.

\bibitem[Fazli et~al.(2013)Fazli, Keshavarzi, and
  Setayeshi]{fazli2013designing}
M.~S. Fazli, K.~Keshavarzi, and S.~Setayeshi, ``Designing a hybrid neuro-fuzzy
  system for classifying the complex data, application on cornea transplant,''
  in \emph{Proceedings on the International Conference on Artificial
  Intelligence (ICAI)}.\hskip 1em plus 0.5em minus 0.4em\relax The Steering
  Committee of The World Congress in Computer Science, Computer Engineering and
  Applied Computing (WorldComp), 2013, p.~1.

\end{thebibliography}

\end{document}